\documentclass[sigconf]{acmart}

\copyrightyear{2026}
\acmYear{2026}
\setcopyright{cc}
\setcctype{by}
\acmConference[MSR '26]{23rd International Conference on Mining Software Repositories}{April 13--14, 2026}{Rio de Janeiro, Brazil}
\acmBooktitle{23rd International Conference on Mining Software Repositories (MSR '26), April 13--14, 2026, Rio de Janeiro, Brazil}
\acmDOI{10.1145/3793302.3793562}
\acmISBN{979-8-4007-2474-9/2026/04}

\AtBeginDocument{%
  }

\usepackage{enumitem}
\usepackage{pifont}
\usepackage{makecell}
\usepackage{multirow}
\usepackage{balance}

\ccsdesc[500]{Software and its engineering~Software libraries and repositories}
\ccsdesc[500]{Computing methodologies~Intelligent agents}

\keywords{LLMs, AI coding agents, software libraries, dependency management, pull requests, empirical software engineering}

\begin{document}

\title{A Study of Library Usage in Agent-Authored Pull Requests}

\author{Lukas Twist}
\orcid{0009-0009-6640-2532}
\affiliation{%
  \institution{King's College London}
  \city{London}
  \country{UK}
}
\email{lukas.twist@kcl.ac.uk}

\author{Jie M. Zhang}
\orcid{0000-0003-0481-7264}
\affiliation{%
  \institution{King's College London}
  \city{London}
  \country{UK}
}
\email{jie.zhang@kcl.ac.uk}

\renewcommand{\shortauthors}{Twist and Zhang}

\begin{abstract}
Coding agents are becoming increasingly capable of completing end-to-end software engineering workflows that previously required a human developer, including raising pull requests (PRs) to propose their changes.
However, we still know little about how these agents use libraries when generating code, a core part of real-world software development.
To fill this gap, we study 26,760 agent-authored PRs from the \texttt{AIDev} dataset to examine three questions: how often do agents import libraries, how often do they introduce new dependencies (and with what versioning), and which specific libraries do they choose?
We find that agents often import libraries (29.5\% of PRs) but rarely add new dependencies (1.3\% of PRs); and when they do, they follow strong versioning practices (75.0\% specify a version), an improvement on direct LLM usage where versions are rarely mentioned. 
Generally, agents draw from a surprisingly diverse set of external libraries, contrasting with the limited ``library preferences'' seen in prior non-agentic LLM studies. 
Our findings offer an early empirical view on how AI coding agents interact with today’s software ecosystems.
\end{abstract}

\maketitle


\section{Introduction}\label{sec:intro}

Large language models (LLMs) have rapidly reshaped the way developers write code, moving beyond single-line completions to generating whole functions or creating new projects~\cite{chenEvaluatingLargeLanguage2021,jiangSurveyLargeLanguage2024}.
More recently, the field has seen a shift from human-centric code completion to \textit{agentic} workflows in which LLMs plan, act, and call tools to solve more realistic software engineering tasks~\cite{liuLargeLanguageModelBased2025}.
These agentic systems---now widely available in production tools~\cite{shrivastavaComparativeFeaturesetAnalysis2025}---can complete end-to-end workflows that previously required a human developer.
Notably, these workflows include raising pull requests (PRs), allowing agents to propose more substantial code changes than earlier bots~\cite{heAutomatingDependencyUpdates2023a}.
Producing such substantial changes requires agents to engage with the same software ecosystems and abstractions relied upon by human developers.

Libraries are central to modern software engineering~\cite{somervilleSoftwareEngineeringGlobal2016}, where reusing established components improves developer productivity and avoids reinventing well-tested functionality~\cite{xuWhyReinventingWheels2020}.
For agents to contribute meaningfully to real-world projects, they must therefore be able to use libraries effectively.
However, prior work shows that LLMs in non-agentic settings often struggle in this respect: exhibiting limited diversity in their library choices~\cite{twist-arxiv-llm-code-bias}, omitting version constraints~\cite{rajRoleLibraryVersions2024a}, or calling deprecated APIs~\cite{wangLLMsMeetLibrary2025}.
To mitigate these issues, existing approaches frequently rely on retrieval-based mechanisms~\cite{taoRetrievalAugmentedCodeGeneration2025, wangExploraCoderAdvancingCode2025}.
Agentic systems, however, have greater autonomy and may bypass such additional infrastructure.
What remains unclear is whether agentic workflows will actually use external libraries---and if they do, whether this autonomy helps mitigate existing problems or simply reproduces them at scale.

To begin to fill this gap, we present the first focused empirical study of how AI coding agents use libraries in practice, by mining real agent-authored PRs from the \texttt{AIDev} dataset~\cite{liRiseAITeammates2025}.
Specifically, we analyse file diffs (both code and dependency manifests) from 26,760 PRs across four languages (\texttt{TypeScript}, \texttt{Python}, \texttt{Go}, and \texttt{C\#}), to answer three research questions (RQs):

\textbf{\textsc{RQ1 Library Usage:}} \textit{How often do agents import libraries when generating code, and do they favour built-in or external libraries?}

\textbf{\textsc{RQ2 New Dependencies:}} \textit{How often do agents introduce new dependencies into a repository, and do they specify versions?}

\textbf{\textsc{RQ3 Choosing Libraries:}} \textit{Which specific libraries do agents choose, and what functional roles do they play?}

Our findings are concise but instructive.
\textbf{\textit{1)}} Agents often import libraries (29.5\% of PRs) but rarely add entirely new dependencies (1.3\% of PRs).
\textbf{\textit{2)}} Agentic workflows encourage far better version hygiene than traditional LLM prompting: agents specify versions for 75.0\% of added dependencies, whereas LLMs would only mention versions in $\approx$9\% of library-related interactions~\cite{rajRoleLibraryVersions2024a}.
\textbf{\textit{3)}} Agents use a diverse range of libraries, importing 3,988 distinct external packages across 7,888 PRs---a sharp contrast to the narrow ``library preferences'' previously observed in more self-contained LLM tasks~\cite{twist-arxiv-llm-code-bias}.

\noindent\textit{Our contributions are as follows.}

\begin{enumerate}
    \item We conduct the first empirical study of agentic library usage by mining 26,760 agent-authored PRs across four languages.
    \item We find that agents often import libraries but rarely introduce new dependencies, and when they do, their versioning practices are unexpectedly strong.
    \item We release our dataset processing and analysis code to support reproducibility and future research in this area: \\
    \href{https://github.com/itsluketwist/agent-library-usage/}{\textit{\textcolor{purple}{https://github.com/itsluketwist/agent-library-usage/}}}
\end{enumerate}

\begin{table*}[ht]
    \caption{
        \textbf{\textit{Language-specific details for data extraction.}}
        This table summarises the code files, dependency manifests, and representative import/dependency syntax used to identify library usage across all four languages.
    }
    \label{tab:experiment}

    \Description{
        Language-specific details used for data extraction, including code file extensions, dependency manifest files, and representative import and dependency syntax for \texttt{TypeScript}, \texttt{Python}, \texttt{Go}, and \texttt{C\#}.
    }

    \centering
    \resizebox{0.98\textwidth}{!}{

    \bgroup
    \def\arraystretch{1.1}
    
    \begin{tabular}{|c|c|c|c|c|}
    \hline

        \thead{Programming\\Language} & 
        \thead{Code File\\Extensions} &
        \thead{Example Import\\Statements} &
        \thead{Dependency\\Manifests} &
        \thead{Example\\Dependencies}
        \\
        
    \hline
        \textbf{\texttt{TypeScript}}~\cite{typescriptDocumentationModules2025} &
        \makecell[l]{.ts, .tsx, .js,\\.jsx, .mjs, .cjs} &
        \makecell[l]{\texttt{import \{ useEffect \} from 'react';}\\\texttt{import \{ Class \} from '@angular/core';}} &
        \makecell[l]{package.json} &
        \makecell[l]{\texttt{"lodash": "$\sim$4.17.21",}}
        \\
    \hline
        \multirowcell{2}{\textbf{\texttt{Python}}~\cite{python5ImportSystem2025}} &
        \multirowcell{2}[0pt][l]{.py} &
        \multirowcell{2}[0pt][l]{\texttt{import numpy as np}\\\texttt{from django.db import models}} &
        \makecell[l]{requirements.txt} &
        \makecell[l]{\texttt{requests}\\\texttt{pandas<2.0}}
        \\
    \cline{4-5}
        & & &
        \makecell[l]{pyproject.toml} &
        \makecell[l]{\texttt{"pytest"}\\\texttt{"numpy==2.79",}}
        \\
    \hline
        \textbf{\texttt{Go}}~\cite{goPackagesGoProgramming2025} &
        \makecell[l]{.go} &
        \makecell[l]{\texttt{import "fmt"}\\\texttt{import ( mux "github.com/gorilla/mux" )}} &
        \makecell[l]{go.mod} &
        \makecell[l]{\texttt{require github.com/pkg/errors v0.9.1}}
        \\
    \hline
        \multirowcell{2}{\textbf{\texttt{C\#}}~\cite{microsoftUsingDirectiveImport2025} } &
        \multirowcell{2}[0pt][l]{.cs} &
        \multirowcell{2}[0pt][l]{\texttt{using Newtonsoft.Json;}\\\texttt{using Serilog;}} &
        \makecell[l]{\{name\}.csproj\\packages.config} &
        \makecell[l]{\texttt{<PackageReference}\\\hspace*{20pt}\texttt{Include="Serilog" Version="2.12.0" />}}
        \\
    \cline{4-5}
        & & &
        \makecell[l]{paket.dependencies} &
        \makecell[l]{\texttt{nuget Newtonsoft.Json 13.0.1}}
        \\
    \hline
    
    \end{tabular}

    \egroup
    
    }
\end{table*}


\section{Dataset}

To conduct our study, we use the \texttt{AIDev} dataset~\cite{liRiseAITeammates2025}, a large-scale corpus of almost a million pull requests (PRs) from over 116,000 GitHub repositories, authored by five distinct LLM-based coding agents (Claude Code, Cursor, Devin, Copilot, OpenAI Codex).
Specifically, we use the November 2025 snapshot (commit \texttt{eee0408}~\cite{hao-liHaoliAIDevEee0408a277826d88fc0ca5fa07d2fc325c96af12025}).

For our analysis, we use the \texttt{AIDev-Pop} subset because it includes the full commit details per PR; \texttt{AIDev-Pop} contains 33,596 PRs from 2,807 high-quality repositories (those with over 100 GitHub stars).
We use three tables: \textit{repository}, containing metadata about each repository; \textit{pull\_request}, which provides high-level PR information; and \textit{pr\_commit\_details}, which contains file-level diffs and commit content.
The dataset provides diffs in the ``combined diff format’’~\cite{gitGitGitdiffDocumentation}, allowing us to extract new content from each file by selecting lines that begin with whitespace followed by exactly one ``+’’.

We restrict our study to the four most common languages in the dataset---those with the largest number of repositories---to ensure both depth of analysis and sufficient statistical support.
The chosen languages are \texttt{TypeScript}/\texttt{JavaScript} (combined due to nearly identical syntax and library ecosystems, referred to as \texttt{TypeScript} throughout the paper), \texttt{Python}, \texttt{Go} and \texttt{C\#}.
We initially included \texttt{Rust} (fifth most popular language), but its PRs relied heavily on internal workspace crates that were indistinguishable from external imports, so we limited our study to the top four languages.

\paragraph{Ethical considerations.}
\texttt{AIDev} contains no personal or sensitive information, and all data is publicly accessible under the open-source license of its respective repository (typically MIT, Apache-2.0, or GPL variants).
As our analysis only uses publicly available code, and does not involve any human-subject data, our use of the dataset does not raise privacy or ethical concerns.


\begin{table*}[ht]
    \caption{
        \textbf{\textit{Coding agent library usage (RQ1, RQ2).}}
        Summary statistics from the analysis of 26,760 agent-authored pull requests, showing how often agents import libraries and introduce new project dependencies.
    }
    \label{tab:results-one}

    \Description{
        Summary statistics for 26,760 agent-authored pull requests, showing per-language counts and percentages for library imports, standard versus external library usage, new dependency additions, version specification, and pull request merge rates.
    }

    \centering
    \resizebox{0.89\textwidth}{!}{

    \bgroup
    \def\arraystretch{1.1}

\begin{tabular}{@{}c|p{5.2cm}|r|r|r|r|r|}
\cline{2-7}
  & \textbf{Metric} & \multicolumn{1}{c|}{\textbf{\texttt{TypeScript}}} & \multicolumn{1}{c|}{\textbf{\texttt{Python}}} & \multicolumn{1}{c|}{\textbf{\texttt{Go}}} & \multicolumn{1}{c|}{\textbf{\texttt{C\#}}} & \textbf{All languages} \\
\hline
\multicolumn{1}{|c|}{\multirow{3}{*}{\makecell{\textbf{General} \\ \textbf{statistics}}}} & Total repositories & 840 & 530 & 242 & 220 & 1,832 \\
\multicolumn{1}{|c|}{} & Total pull requests & 7,480 & 7,190 & 10,107 & 1,983 & 26,760 \\
\multicolumn{1}{|c|}{} & Total PRs merged & 4,428 (59.2\%) & 5,468 (76.1\%) & 8,361 (82.7\%) & 1,146 (57.8\%) & 19,403 (72.5\%) \\
\hline
\multicolumn{1}{|c|}{\multirow{7}{*}{\makecell{\textbf{RQ1 \textsc{Library}} \\ \textbf{\textsc{Usage}}}}} & PRs importing any library & 2,993 (40.0\%) & 2,720 (37.8\%) & 1,273 (12.6\%) & 902 (45.5\%) & 7,888 (29.5\%) \\
\multicolumn{1}{|c|}{} & PRs importing standard library & 828 (11.1\%) & 2,383 (33.1\%) & 1,244 (12.3\%) & 743 (37.5\%) & 5,198 (19.4\%) \\
\multicolumn{1}{|c|}{} & PRs importing external libraries & 2,847 (38.1\%) & 1,914 (26.6\%) & 324 (3.2\%) & 738 (37.2\%) & 5,823 (21.8\%) \\
\multicolumn{1}{|c|}{} & Avg. standard library imports per PR & 0.26 & 1.28 & 0.63 & 0.49 & 0.69 \\
\multicolumn{1}{|c|}{} & Avg. external library imports per PR & 1.87 & 0.73 & 0.08 & 0.73 & 0.80 \\
\multicolumn{1}{|c|}{} & Unique external libraries imported & 2,495 & 948 & 314 & 231 & 3,988 \\
\multicolumn{1}{|c|}{} & PRs with library imports merged & 1,595 (53.3\%) & 1,768 (65.0\%) & 918 (72.1\%) & 467 (51.8\%) & 4,748 (60.2\%) \\
\hline
\multicolumn{1}{|c|}{\multirow{6}{*}{\makecell{\textbf{RQ2 \textsc{New}} \\ \textbf{\textsc{Dependencies}}}}} & PRs adding new dependencies & 9 (0.1\%) & 200 (2.8\%) & 16 (0.2\%) & 120 (6.1\%) & 345 (1.3\%) \\
\multicolumn{1}{|c|}{} & Avg. new dependencies per PR & 0.01 & 0.17 & 0.02 & 0.30 & 0.08 \\
\multicolumn{1}{|c|}{} & Total new dependencies & 95 & 1,237 & 203 & 593 & 2,128 \\
\multicolumn{1}{|c|}{} & Dependencies specifying version & 95 (100.0\%) & 1,020 (82.5\%) & 203 (100.0\%) & 278 (46.9\%) & 1,596 (75.0\%) \\
\multicolumn{1}{|c|}{} & Unique new dependencies & 85 & 552 & 188 & 294 & 1,119 \\
\multicolumn{1}{|c|}{} & PRs with new dependencies merged & 6 (66.7\%) & 131 (65.5\%) & 5 (31.2\%) & 56 (46.7\%) & 198 (57.4\%) \\
\hline
\end{tabular}

    \egroup

    }
\end{table*}

\section{Data Extraction Method} \label{sec:experiment}

In this section, we outline how we extract the data used to answer each RQ.
Full implementation details--including all parsing scripts and processing steps--are available in our public GitHub repository.

\subsection{Method for \textsc{RQ1 Library Usage}}

We begin by investigating how inclined agents are to make use of libraries in their code.
We recognise library usage only when imports appear in commits, because imports that already existed in a file prior to the agent’s PR do not reflect the agent’s preferences; and with only partial file snapshots provided in diffs, downstream function calls are often ambiguous without the corresponding import statement.
We use file extensions to identify code files for each programming language, as defined by GitHub Linguist~\cite{github-linguistLinguistLanguageSavant}.

We implement language-specific \texttt{regex} parsers to automatically extract import statements from those files, taking care to distinguish relative imports (from within the project itself).
All extraction is supported by unit tests to ensure reliability.
Information on the file extensions we consider, with example imports, is provided in Table~\ref{tab:experiment}.
Each library is categorised as either standard (built-in, defined in the language documentation) or external (installed via a third-party package manager), allowing us to measure how much agents rely on external ecosystems versus native language capabilities.

\subsection{Method for \textsc{RQ2 New Dependencies}}

Next, we examine how often agents introduce new dependencies into repositories.
Here, we are particularly interested in the quality of these additions--specifically, whether agents are explicit with version constraints--because prior work shows that LLMs in non-agentic workflows rarely discuss versions~\cite{rajRoleLibraryVersions2024a}, despite the known risks.
For each language, we identify its corresponding dependency manifest files (for example, ``requirements.txt'' for \texttt{Python}, ``package.json'' for \texttt{TypeScript}) following the GitHub Docs~\cite{githubDependencyGraphSupported}; we ignore lock files, which contain large numbers of second-hand dependencies not explicitly used within the repository.

We implement manifest-specific \texttt{regex} parsers to automatically identify new dependencies--including any explicit version specifiers (for example, ``==1.2'' or ``>=1.0'')--again with unit tests to ensure reliability.
The manifest files we consider, with example dependencies, are listed in Table~\ref{tab:experiment}.
This enables us to quantify how often agents introduce new dependencies and how reliably they do so.

\subsection{Method for \textsc{RQ3 Choosing Libraries}}

Finally, we examine the specific libraries agents choose and the functional roles they play.
For each language, we take the set of libraries identified in \textsc{RQ1} and \textsc{RQ2} and compute simple frequency statistics to surface the most commonly used or newly introduced libraries.
We then perform manual analysis on the top ten libraries for each language and RQ to determine the areas that agents are most likely to use external libraries.

\section{Findings \& Discussion}\label{sec:results}

We analyse a total of 26,760 agent-authored pull requests across 1,832 repositories.
Here, we present and discuss our findings.

\subsection{Findings for \textsc{RQ1 Library Usage}}

For this RQ, we extract imports from agent-authored code to investigate their library usage patterns.
\textit{Results in Table~\ref{tab:results-one}.}

Agents often import libraries when generating code, with 29.5\% of PRs importing at least one library, and an average of 1.5 libraries imported across all PRs.
However, the extent varies substantially by language: agents lean most heavily on libraries when working in \texttt{TypeScript} and \texttt{C\#} (40.0\% and 45.5\% of PRs), while \texttt{Go} is a clear outlier, importing libraries in just 12.6\% of PRs.
This suggests that agents primarily perform localised modifications in \texttt{Go}, potentially reflecting both the language’s strong standard library and ecosystem norms that discourage heavy external dependencies.
This pattern appears again when considering average imports per PR: both \texttt{TypeScript} and \texttt{Python} exceed two libraries per PR, whereas \texttt{Go} averages below one.
Although reusing libraries is a core skill for professional developers, there is no established baseline for how often this should occur, making it difficult to assess whether these rates indicate overuse, underuse, or neither.
Despite these substantial cross-language differences, PRs that import libraries are merged at rates only 6--11\% lower than PRs that do not, suggesting that library usage itself is not strongly penalised during review.

Agents rely strongly on the standard library in \texttt{Python} and \texttt{C\#} (33.1\% and 37.5\% of PRs), while their use of external libraries in those languages remains comparatively balanced (26.6\% and 37.2\%).
In \texttt{TypeScript}, the pattern is very different: agents overwhelmingly favour external packages (38.1\% of PRs) over the standard library (11.1\%), indicating a strong preference for high-level tooling.
In contrast, \texttt{Go} uses standard libraries almost four times more often than external libraries.
Across all languages, agents collectively import 3,988 unique external libraries, indicating substantial diversity.
This contrasts sharply with work showing that LLMs typically exhibit low library diversity~\cite{twist-arxiv-llm-code-bias}, suggesting that rich project-level context and greater autonomy encourages broader library use.
While it may seem surprising that \texttt{TypeScript} alone contributes more than half of all unique libraries, this aligns with the ecosystem’s well-known fragmentation and reliance on micro-packages~\cite{kulaImpactMicroPackagesEmpirical2017}; however, such breadth may also increase the risk of inconsistent dependency choices across PRs.

\textbf{RQ1 Summary.}
Agents often import libraries for \texttt{TypeScript} (40.0\% of PRs), \texttt{Python} (37.8\%) and \texttt{C\#} (45.5\%), but rarely for \texttt{Go} (12.6\%).
For \texttt{TypeScript}, agents favour external libraries over built-in functionality; for \texttt{Python} and \texttt{C\#}, library usage is more balanced; and for \texttt{Go}, the standard library is strongly preferred.


\begin{table*}[ht]
    \caption{
        \textbf{\textit{Libraries used by agents (RQ3).}}
        Top external libraries imported and top newly added dependencies by language, with number of PRs. The lists reveal the functional roles (testing, scaffolding, UI, data, API clients) that agents most commonly serve.
    }
    \label{tab:results-two}

    \Description{
        Lists of the ten most frequently imported external libraries and the ten most frequently added new dependencies for each language, along with the number of pull requests in which each library appears.
    }

    \centering
    \resizebox{0.98\textwidth}{!}{

    \bgroup
    \def\arraystretch{1.1}

\begin{tabular}{|l|l|l|l|l|}
\hline
\textbf{Metric} & \multicolumn{1}{c|}{\textbf{\texttt{TypeScript}}} & \multicolumn{1}{c|}{\textbf{\texttt{Python}}} & \multicolumn{1}{c|}{\textbf{\texttt{Go}}} & \multicolumn{1}{c|}{\textbf{\texttt{C\#}}} \\
\hline
\makecell[l]{Top 10 external \\ library imports} & \makecell[l]{react (1013) \\ vitest (483) \\ zod (313) \\ next (303) \\ lucide-react (261) \\ @testing-library/react (136) \\ @tanstack/react-query (111) \\ lodash (107) \\ drizzle-orm (97) \\ chai (92)} & \makecell[l]{pytest (812) \\ pydantic (279) \\ crewai (264) \\ requests (125) \\ numpy (125) \\ fastapi (114) \\ openai (65) \\ pandas (60) \\ httpx (53) \\ yaml (52)} & \makecell[l]{github.com/stretchr/testify (108) \\ github.com/smacker/go-tree-sitter (21) \\ github.com/spf13/cobra (20) \\ github.com/cometbft/cometbft (13) \\ github.com/microsoft/typescript-go (13) \\ github.com/tliron/glsp (13) \\ go.uber.org/zap (12) \\ github.com/gofiber/fiber (11) \\ k8s.io/apimachinery/pkg (10) \\ github.com/onsi/gomega (9)} & \makecell[l]{Avalonia (275) \\ Xunit (243) \\ Dock (131) \\ Aspire (55) \\ Azure (50) \\ NUnit (48) \\ FluentAssertions (30) \\ ReactiveUI (28) \\ Newtonsoft (27) \\ Svg (27)} \\
\hline
\makecell[l]{Top 10 new \\ dependencies} & \makecell[l]{typescript (3) \\ vite (3) \\ husky (2) \\ @biomejs/biome (2) \\ zod (2) \\ prettier (2) \\ @types/node (2) \\ three (2) \\ @electron-forge/plugin-auto-unpack-natives (1) \\ eslint-plugin-react-compiler (1)} & \makecell[l]{requests (24) \\ openai-agents (22) \\ openai (21) \\ numpy (21) \\ pandas (18) \\ pytest (14) \\ python-dotenv (13) \\ uvicorn (13) \\ pydantic (12) \\ aiohttp (12)} & \makecell[l]{github.com/golang/snappy (6) \\ github.com/stretchr/testify (3) \\ gopkg.in/yaml.v3 (3) \\ github.com/klauspost/reedsolomon (3) \\ github.com/gogo/protobuf (3) \\ go.uber.org/atomic (2) \\ github.com/smacker/go-tree-sitter (2) \\ github.com/leonklingele/grouper (1) \\ github.com/sashamelentyev/usestdlibvars (1) \\ github.com/breml/bidichk (1)} & \makecell[l]{Microsoft.NET.Test.Sdk (26) \\ xunit.runner.visualstudio (22) \\ xunit (20) \\ Moq (20) \\ coverlet.collector (15) \\ Microsoft.AspNetCore.Mvc.Testing (11) \\ NUnit (9) \\ NUnit3TestAdapter (9) \\ Azure.Identity (7) \\ Microsoft.Extensions.DependencyInjection (7)} \\
\hline
\end{tabular}

    \egroup

    }
\end{table*}

\subsection{Findings for \textsc{RQ2 New Dependencies}}

For this RQ, we examine dependency manifests to investigate how often agents add new dependencies to projects.
\textit{Results in Table~\ref{tab:results-one}.}

Across all languages, dependency addition is extremely rare: only 1.3\% of PRs introduce even a single new library.
\texttt{C\#} and \texttt{Python} are the only languages where dependency addition is non-negligible (6.1\% and 2.8\% of PRs).
This conservative behaviour contrasts with typical human workflows in large ecosystems---such as \texttt{TypeScript} and \texttt{Python}---where adding a dependency to reuse existing functionality is common~\cite{abdalkareemImpactUsingTrivial2020}.
Agents often struggle with installing packages and resolving environment conflicts~\cite{aroraSetupBenchAssessingSoftware2025}, suggesting that their reluctance to add dependencies may reflect practical limitations rather than deliberate conservatism.
Additionally, the presence of new dependencies is associated with a substantially lower merge rate only for \texttt{Go} PRs, where the merge rate drops by over 50\%, compared to reductions of at most 10\% in other languages.

When agents do introduce new dependencies, they generally do so with good versioning practices.
Across all languages, 75.0\% of new dependencies include an explicit version constraint, with \texttt{TypeScript} and \texttt{Go} specifying versions in every instance.
This contrasts with previous, non-agentic, LLM research that observes version constraints mentioned in only $\approx$9\% of library-related discussions~\cite{rajRoleLibraryVersions2024a}.
Whilst not directly comparable, this does indicate that agentic setups may support more robust dependency management over direct LLM usage; likely because agents benefit from full project context and direct interaction with dependency manifests.

Finally, we observe substantial diversity in the dependencies added: 1,119 unique libraries across 2,128 total additions.
This high ratio of unique packages suggests that agents do not default to a small, preferred set of libraries when adding new dependencies.
Much of this diversity comes from a small number of diverse PRs that create entirely new projects and therefore introduce many dependencies at once; a pattern consistent with known strengths of modern LLMs in scaffolding initial project structures~\cite{rasnayakaEmpiricalStudyUsage2024}.

\textbf{RQ2 Summary.}
Agents rarely add new dependencies (1.3\% of PRs), most commonly for \texttt{C\#} (6.1\%) and \texttt{Python} (2.8\%).
When they do add dependencies, they have strong version hygiene, specifying a constraint in 75.0\% of cases.

\subsection{Findings for \textsc{RQ3 Choosing Libraries}}

For this RQ, we analyse trends in the libraries that agents most commonly use.
\textit{Table~\ref{tab:results-two} lists the most common libraries.}

Across all languages, test frameworks and development-tooling libraries are widely used, but they make up only a minority of the distinct most-imported libraries (10--40\%), with most imports instead being production-orientated.
In contrast, newly added dependencies skew heavily toward testing and tooling, indicating that when agents modify dependency manifests, they prioritise scaffolding and validation over expanding runtime dependency graphs--a conservative and security-conscious pattern.

Library choices also align closely with language ecosystem norms.
\texttt{TypeScript} imports focus on the front-end stack (\texttt{react}, \texttt{next}, \texttt{lucide-react}).
\texttt{Python} shows a strong AI/data-science presence (\texttt{crewai}, \texttt{openai}, \texttt{numpy}, \texttt{pandas}) alongside various HTTP clients.
\texttt{Go} leans toward infrastructure and cloud-native tooling (kubernetes packages, logging, CLI frameworks) alongside parser/protocol tools.
\texttt{C\#} imports include desktop/UI frameworks (\texttt{Avalonia}, \texttt{Dock}) other Microsoft-ecosystem libraries, while new dependencies disproportionately add test infrastructure (\texttt{xUnit}, \texttt{NUnit}, \texttt{Moq}).
These results suggest that agentic workflows do not select libraries arbitrarily, but instead exhibit ecosystem-aware and risk-conscious behaviour when choosing both production libraries and new dependencies.
\textbf{RQ3 Summary.}
Across languages, agents import a broad set of production libraries but predominantly add testing and developer tooling dependencies, reflecting a conservative, ecosystem-aware approach to dependency growth.

\section{Threats to Validity}

First, our results reflect behaviour in only a subset of agent-authored PRs, covering five deployed AI coding agents, four programming languages, and repositories with 100+ stars; other agents, languages, or repositories may exhibit different patterns.
For internal validity, our extraction relies on \texttt{regex} parsing of file diffs.
This is the most practical choice because abstract syntax tree parsing is not feasible on partial diffs, but may miss edge cases due to incomplete context or complex syntax.
We mitigate this threat through manual validation of 100 diffs per language and comprehensive unit testing.

\section{Conclusion}\label{sec:conclusion}

We presented the first empirical study of how AI coding agents use libraries, analysing 26,760 agent-authored PRs across four languages.
Our results show that agents frequently import libraries (29.5\% of PRs) but rarely add new dependencies (1.3\% of PRs), and when they do, versioning practices are strong.
Agents also demonstrate surprisingly broad library diversity, with library imports centred around production code and new dependencies focussed on test code and developer tooling.

Looking ahead, the next step is to probe not just \textit{which} libraries agents use, but \textit{how well} they use them.
For example, do they reproduce known LLM failure modes, such as hallucinating library names~\cite{twistLibraryHallucinationsLLMs2025} or relying on deprecated features~\cite{wangLLMsMeetLibrary2025}?
A deeper analysis of the code they generate would also help establish whether agents sensibly reuse existing libraries or quietly reimplement functionality.
As agentic workflows continue to mature, understanding these behaviours will be key to integrating them safely and effectively into real software development practices.

\begin{acks}
    This work was supported by the UKRI Centre for Doctoral Training in Safe and Trusted AI (EPSRC iCASE Award ref EP/Y528572/1).
\end{acks}

\newpage

\balance

\bibliographystyle{ACM-Reference-Format}
\bibliography{main}

\end{document}